\documentclass{ws-p10x7}
\usepackage{axodraw}
\newcommand{\spqr}{SPQ$_{\textrm{CD}}$R\ }
\begin{document}

\title{Phenomenology from Lattice QCD}

\author{C. T. Sachrajda}

\address{Department of Physics and Astronomy, University of
Southampton,\\ Southampton SO17~1BJ, UK
\\E-mail: cts@hep.phys.soton.ac.uk}

\twocolumn[\maketitle\abstract{I review two subjects in which
lattice simulations are making, or can make in the future, a
significant contribution to particle physics phenomenology. The
first subject is the evaluation of quantities which enter into the
determination of the vertex $A$ of the unitarity triangle from
experimental measurements of decay rates and mixing amplitudes.
These quantities include the form-factors for semileptonic
$B$-decays, the leptonic decay constants of the $B$ and $B_s$
mesons and the $B$-parameters for $B_{d}$-$\bar B_{d}$ and
$B_{s}$-$\bar B_{s}$ mixing and the corresponding parameter for
$K$-$\bar K$ mixing, $B_K$. In the second part of this talk I will
review the status and prospects for the evaluation of $K\to\pi\pi$
amplitudes and for the subsequent study of the $\Delta I=1/2$ rule
and the evaluation of $\varepsilon^\prime/\varepsilon$.}]

\section{Introduction}

In this lecture I will briefly review some of the contributions
which lattice computations are making to particle physics
phenomenology. The lattice formulation of quantum field theory
together with large-scale numerical simulations is contributing to
a wide variety of fundamental questions in particle physics, both
theoretical and phenomenological. Here I will concentrate on one
of the most important r\^oles of lattice QCD, the evaluation of
non-perturbative QCD effects in physical amplitudes and other
quantities. Indeed, it is frequently our inability to quantify the
long-distance QCD effects in weak processes which is the dominant
source of uncertainty in determining fundamental quantities from
experimental measurements and lattice simulations provide an {\it
ab initio} framework for the evaluation of these effects.

For many physical quantities lattice calculations have been
performed for over ten years and the emphasis is on the reduction
of systematic uncertainties. I will briefly outline the sources of
some of the other uncertainties when presenting the results below,
but let me now mention two important sources of error which the
community is striving to reduce. The first of these is
\textit{quenching}, the neglect of vacuum polarization and other
quark loop-effects. Most large-scale phenomenological calculations
have been performed in the quenched approximation, although
increasingly calculations are being performed with 2 flavours of
sea quarks. Although it is natural to be skeptical about quenched
calculations and consider full QCD (unquenched) ones as totally
reliable, in my opinion neither of these views is fully justified
at present. Where results from quenched calculations can be
compared with experimental measurements, they typically agree to
within 10\% or so, which is sufficiently accurate for some
quantities and not so for others. On the other hand, it should
also be remembered that in unquenched simulations the masses of
the sea and valence quarks are large ($m_\pi/m_\rho$ is typically
about 0.6 or more) so that significant extrapolations are needed,
and this also leads to uncertainties. This brings me to the second
source of lattice systematic error, which is increasingly being
studied in detail, the extrapolation to the {\it chiral limit}. In
order to avoid unphysical effects due to the finite volume of the
lattice, simulations are performed with up and down quarks with
masses ($m_u$ and $m_d$) in the region of $m_s$, the mass of the
strange quark, and the results are then extrapolated to the
physical values of $m_u$ and $m_d$ (the computing cost also
increases dramatically as the masses of the quarks decrease). For
small values of the quark masses we can hope to exploit chiral
symmetry (and chiral perturbation theory, in particular) to guide
us in this extrapolation. The main question is therefore whether
there is a region of overlap between the range of masses being
used in simulations and those which are sufficiently light for
chiral perturbation theory to apply. An added subtlety is that the
chiral structure of the quenched theory is very different from
full QCD. I will not describe these studies further, but they
represent an important step towards the improvement of the
reliability and precision of lattice computations.

I would like to stress that the material presented here represents
only a small fraction of lattice results in general and lattice
phenomenology in particular. The proceedings of the annual lattice
conferences~\cite{latts} contain detailed reviews of lattice
contributions to different areas of phenomenology, as well as
original contributions from the groups carrying out the studies.
For a summary of the future prospects in the subject I refer you
to the report of an ECFA panel which was charged with the task of
considering these~\cite{ECFA}.

In this talk I will focus on two topics, the contribution that
lattice simulations are making to the determination of the vertex
$A$ of the unitarity triangle (section~\ref{sec:ckm}) and the
status of lattice computations of the amplitudes for $K\to\pi\pi$
decays and the prospect for the improvement in the precision of
these calculations (section~\ref{sec:kpipi}).

The annual International Symposia on Lattice Field Theory provide
an important forum for the collaborations to present their new
results. This lecture was delivered before the 2001 Lattice
conference (Latt2001), and in consequence the results presented in
this written version were also largely compiled before Latt2001.
However, the new results presented for kaon decays at Latt2001,
and for $\varepsilon^\prime/\varepsilon$ in particular, are of
considerable interest and I felt that it was necessary to present
and discuss them. This is done in section~\ref{subsec:prospects}.

\section{Lattice QCD and the Unitarity
Triangle}\label{sec:ckm}

\begin{table*}[t]
\begin{center}
\caption{Schematic table of some of the experimental quantities which are
measured from which information about the unitarity triangle is
deduced (from ref.~\protect\cite{stocchi}). Factors in bold-type
represent quantities which
are calculated in lattice computations.} \label{tab:latinputs}
\begin{tabular}{|c|c|c|}\hline
Measurement&$V_{\textrm{CKM}}\times$ Other&Constraint\\ \hline
$b\to u/b\to c$&$|V_{ub}/V_{cb}|^2$&$\bar \rho^2+\bar\eta^2$\\
$\Delta m_d$&$|V_{td}|^2${\boldmath $f_{B_d}^2B_{B_d}$}$f(m_t)$&
$(1-\bar \rho)^2+\bar\eta^2$\\
$\Delta m_d/\Delta m_s$&
$|V_{td}/V_{ts}|^2${\boldmath $(f_{B_d}^2B_{B_d})/(f_{B_s}^2B_{B_s})$}&
$(1-\bar \rho)^2+\bar\eta^2$\\
$\varepsilon_K$&$f(A,\bar\eta,\bar\rho,${\boldmath $B_K$})&
$\propto\bar\eta(1-\bar\rho)$\\ \hline
\end{tabular}
\end{center}
\end{table*}

In a number of talks at this conference we have seen the current
status of the determination of the vertex $A$ of the unitarity
triangle. Over-determination of the position of $A$ is a
convenient way of testing the consistency of the standard model of
particle physics and of constraining its parameters. Theoretical
inputs, and in particular quantitative estimates of
non-perturbative QCD effects, are required in order to determine
the possible locus of the vertex from measurements of quantities
such as the amplitudes of $K^0-\bar K^0$ mixing or studies of
$B^0-\bar B^0$ mixing~\footnote{The determination of
$\sin(2\beta)$ from the mixing-induced CP-asymmetry in $B\to
J\Psi/ K_S$ is a beautiful and rare exception where there are
essentially no hadronic uncertainties.}. Lattice QCD provides the
opportunity for evaluating these non-perturbative effects, and in
table~\ref{tab:latinputs} (taken from ref.~\cite{stocchi}) I
present a number of the most important examples. The factors in
bold-type in the second column of table~\ref{tab:latinputs} are
quantities which are frequently taken from lattice calculations.
In the first half of this lecture I will review the status of the
determination of these quantities. In addition however, it should
be noted that lattice calculations are also used in the
determination of the CKM matrix elements $V_{ub}$ and $V_{cb}$
from exclusive semileptonic $B$-decays, and I start with a brief
review of these calculations.
\subsection{Exclusive Semi-Leptonic $B$-Decays}

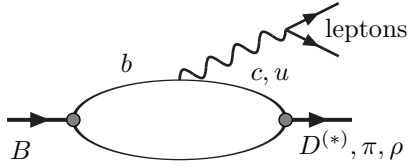
\begin{figure}
\caption{Schematic representation of semileptonic $B$-decays.
\label{fig:sl}}
\begin{center}
\begin{picture}(200,45)(-90,-10)
\Oval(0,0)(15,40)(0)
\SetWidth{1.5}\ArrowLine(-65,0)(-40,0)\ArrowLine(40,0)(65,0)
\SetWidth{1}
\Photon(0,15)(40,34){3}{4}
\ArrowLine(40,34)(60,44)\ArrowLine(40,34)(60,24)
\SetWidth{0.5}
\GCirc(-40,0){2.5}{0.5}\GCirc(40,0){2.5}{0.5}
\Text(-60,-8)[t]{$B$}\Text(45,-4)[lt]{$D^{(\ast)},\pi,\rho$}
\Text(55,34)[l]{leptons} \Text(-20,18)[b]{$b$}
\Text(27,13)[bl]{$c,u$}
\end{picture}
\end{center}
\end{figure}

The CKM matrix elements $V_{cb}$ and $V_{ub}$ are determined from
measurements of inclusive or exclusive semileptonic $B$-decays.
There have been a number of lattice computations of the exclusive
form-factors, which combined with the experimental measurements of
the amplitudes, allows the CKM matrix elements to be determined.
The quark flow diagram for exclusive semi-leptonic $B$ decays is
represented in fig.~\ref{fig:sl}. Lorentz and parity invariance
allows us to write the decay amplitudes in terms of invariant form
factors, e.g. for a decay into a pseudoscalar $P$ ($P=D$ or $\pi$
for example) we can write:
\begin{eqnarray}
\langle P(p_P)|V_\mu(0)|B(p_B)\rangle&=&
\mathbf{f_0(q^2)}\frac{M_B^2-M_P^2}{q^2}q_\mu\nonumber\\
&&\hspace{-1.5in}+
\mathbf{f_+(q^2)}\left[(p_B+p_P)_\mu-\frac{M_B^2-M_P^2}{q^2}q_\mu
\right]\ ,
\end{eqnarray}
where in this case there are two form-factors $f_0$ and $f_+$.
Parity invariance implies that only the vector component $V$ of
the weak $V-A$ current contributes when the final-state hadron is
a pseudoscalar. For $B\to$ vector decays, both the vector and
axial-vector currents contribute and there are four form-factors
($A_{1,2},A(=A_0$-$A_3),V$).

\vspace{5pt}\noindent\underline{{$B\to
D^{(*)}\ell\nu$}\textit{-Decays:}} Lattice~calculations could, in
principle and perhaps also in practice, make a contribution to the
determination of $V_{cb}$ by determining the corresponding form
factors. This is a challenging task however, since, in order to
make an impact, one needs to calculate small corrections to the
result in the heavy quark limit. In particular, for $B\to\rho$
decays we have:
\begin{equation}
\frac{d\Gamma(B\to D^\ast l\nu)}{d\omega}=\textrm{K}
\times |V_{cb}|^2\,{\cal F}^2(\omega)\ ,
\end{equation}
where K is a known kinematic factor, $\omega\equiv v_B\cdot
v_{D^\ast}$ ($v_B$ and $v_{D^\ast}$ are the four-velocities of the
$B$ and $D^\ast$ mesons respectively) and ${\cal F}(1)=1 +
\textrm{corrections}$. In order to make a contribution one has to
be able to calculate the $1/m_Q^2$ non-perturbative corrections to
the distribution at zero recoil ($\omega=1$). There has been a
suggestion~\cite{hashimoto} that by calculating the ratio of
ratios
\begin{equation}
\frac{\langle D|\bar c\gamma_0b|\bar B\rangle\, \langle \bar
B|\bar b\gamma_0c|D\rangle} {\langle D|\bar c\gamma_0c|D\rangle\,
\langle \bar B|\bar b\gamma_0b|\bar B\rangle}
\end{equation}
it may be possible to determine the form-factors sufficiently
accurately. This clearly requires an excellent control of the
systematic errors and it remains to be seen whether sufficient
precision will be possible. I mention in passing that lattice
simulations have been performed which reproduce the CLEO data on
the $\omega$ distribution~\cite{ukqcdiw}. I will however, focus
more on the $b\to u$ decays where the opportunity for lattice
calculations to make a contribution to phenomenological studies is
greater.

\begin{figure}[t]
\begin{center}\caption{Recent data for the form-factors $f_+$ and
$f_0$ of semileptonic $B\to\pi l\nu$ decays from C.~Bernard's
review talk at Lattice
2000~\protect\cite{cblat2000}.\label{fig:f0f+}}
\vspace{-0.4in}\hspace{-0.2in}
\includegraphics[bb = 100  200 4096 4196,
width=2.5truein,height=2.3truein]{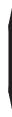}
\end{center}\end{figure}

\vspace{5pt}\noindent \underline{$B\to
\pi,\rho\ell\nu$\textit{-Decays:}} Simulations of $B\to\pi$ or
$\rho$ semileptonic decays yield the form factors at large values
of $q^2$. The reason for this limitation is that the momenta of
the pion or $\rho$-meson must be kept small in order to avoid
artefacts due to the granularity of the lattice. The calculations
have been performed for many years now, and in fig.~\ref{fig:f0f+}
I reproduce a figure from Claude Bernard's review talk at last
year's lattice conference, showing results for the $B\to\pi$ form
factors from four collaborations in the region
19\,GeV$^2<q^2<23$\,GeV$^2$. The collaborations use different
formulations for the $b$-quark, but the results are in reasonable
agreement.

Although much effort is being devoted to extrapolating the lattice
results to smaller values of $q^2$, using as many theoretical
constraints as possible (such as heavy-quark symmetries, unitarity
and analyticity, kinematical constraints and soft-pion relations),
the most meaningful applications of lattice results are (and/or
will be) to the experimental distributions at large values of
$q^2$. A preliminary comparison of the UKQCD lattice data to the
CLEO measurement of the contribution to the width for $B\to\rho$
decays from the region 14\,GeV$^2<q^2<20.3$\,GeV$^2$ (which
corresponds to CLEO's large $q^2$ bin) gives:
\begin{eqnarray}
\Delta\Gamma&=&(7.1^{+1.6}_{-1.0})\,|V_{ub}|^2\,10^{12}\,
\textrm{s}^{-1}\nonumber\\ &&\textrm{\ \
UKQCD~\cite{ukqcd_vub}},\\ &=&(7.1\pm 2.4)\times
10^{7}\,\textrm{s}^{-1}\nonumber\\
&&\hspace{0.2in}\textrm{CLEO~\cite{cleo_vub}},
\end{eqnarray}
from which one obtains $|V_{ub}|=(3.2 \pm 0.6)\times 10^{-3}$. As
the experimental statistics increases and lattice results get more
precise, such comparisons will become the most direct way of
determining $V_{ub}$ from exclusive decays.

\subsection{The Decay Constants $f_B$ and $f_{B_s}$}
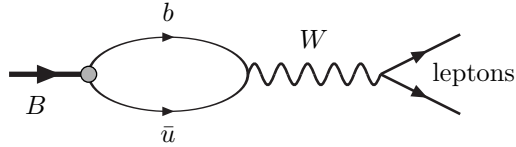
\begin{figure}
\caption{Schematic representation of the leptonic decay of the
$B$-meson.\label{fig:fbdiagram}}
\begin{center}
\begin{picture}(200,40)(-60,-20)
\Text(-50,-8)[t]{$B$} \Text(0,20)[b]{$b$}\Text(0,-20)[t]{$\bar u$}
\Text(55,10)[b]{$W$} \Text(100,0)[l]{leptons}
\ArrowLine(-0.5,14)(0.5,14)\ArrowLine(-0.5,-14)(0.5,-14)
\SetWidth{2}\ArrowLine(-60,0)(-30,0)\SetWidth{0.5}
\Oval(0,0)(14,30)(0)\GCirc(-30,0){3}{0.7} \SetWidth{1.5}
\SetWidth{1}
\Photon(30,0)(80,0){4}{5}\ArrowLine(80,0)(110,15)
\ArrowLine(80,0)(110,-15)
\end{picture}
\end{center}
\end{figure}

\begin{figure}[t]
\caption{Results for $f_B$ from various groups in the quenched
approximation~\protect\cite{fbfigq}. Statistical and systematic
errors have been combined in quadrature.\label{fig:fbquenched}}
\begin{center}
\begin{picture}(200,310)(50,190)
\GBox(155,200)(195,500){0.95}
\Text(120,495)[r]{FNAL97}
\Text(120,475)[r]{APE97}
\Text(120,455)[r]{JLQCD98}
\Text(120,435)[r]{MILC98}
\Text(120,415)[r]{AliKhan98}
\Text(120,395)[r]{JLQCD99}
\Text(120,375)[r]{APE99}
\Text(120,355)[r]{APE00}
\Text(120,335)[r]{UKQCD00}
\Text(120,315)[r]{MILC00}
\Text(120,295)[r]{CPPACS00}
\Text(120,275)[r]{CPPACS00(NR)}
\Text(120,255)[r]{Lellouch-Lin00}
\SetWidth{1}
\Line(164,495)(174,495)\Line(164,495)(153,495)
\GCirc(164,495){1.5}{0.5}
\Line(180,475)(212,475)\Line(180,475)(148,475)
\GCirc(180,475){1.5}{0.5}
\Line(173,455)(177,455)\Line(173,455)(169,455)
\GCirc(173,455){1.5}{0.5}
\Line(157,435)(168,435)\Line(157,435)(146,435)
\GCirc(157 ,435){1.5}{0.5}
\Line(147,415)(158,415)\Line(147,415)(136,415)
\GCirc(147,415){1.5}{0.5}
\Line(167,395)(174,395)\Line(167,395)(160,395)
\GCirc(167,395){1.5}{0.5}
\Line(179,375)(197,375)\Line(179,375)(161,375)
\GCirc(179,375){1.5}{0.5}
\Line(174,355)(196,355)\Line(174,355)(152,355)
\GCirc(174,355){1.5}{0.5}
\Line(218,335)(223,335)\Line(218,335)(213,335)
\GCirc(218,335){1.5}{0.5}
\Line(173,315)(179,315)\Line(173,315)(167,315)
\GCirc(173,315){1.5}{0.5}
\Line(188,295)(191,295)\Line(188,295)(185,295)
\GCirc(188,295){1.5}{0.5}
\Line(191,275)(196,275)\Line(191,275)(186,275)
\GCirc(191,275){1.5}{0.5}
\Line(177,255)(194,255)\Line(177,255)(160,255)
\GCirc(177,255){1.5}{0.5}
\Line(177,455)(185.6,455)\Line(169,455)(160.35,455)
\Line(168,435)(193,435)\Line(146,435)(143,435)
\Line(158,415)(164,415)\Line(136,415)(127,415)
\Line(174,395)(184,395)\Line(160,395)(150,395)
\Line(197,375)(217,375)\Line(161,375)(159,375)
\Line(196,355)(197.4,355)
\Line(223,335)(225,335)\Line(213,335)(177,335)
\Line(179,315)(190,315)\Line(167,315)(156,315)
\Line(191,295)(194,295)\Line(185,295)(182,295)
\Line(196,275)(203,275)\Line(186,275)(179,275)
\Line(194,255)(205,255)\Line(160,255)(146,255)
\Line(50,200)(220,200)\Line(100,200)(100,210)
\Line(150,200)(150,210)\Line(200,200)(200,210)
\Line(60,200)(60,205)\Line(70,200)(70,205)
\Line(80,200)(80,205)\Line(90,200)(90,205)
\Line(100,200)(100,205)\Line(110,200)(110,205)
\Line(120,200)(120,205)\Line(130,200)(130,205)
\Line(140,200)(140,205)\Line(150,200)(150,205)
\Line(160,200)(160,205)\Line(170,200)(170,205)
\Line(180,200)(180,205)\Line(190,200)(190,205)
\Line(200,200)(200,205)\Line(210,200)(210,205)
\Line(220,200)(220,205)
\Text(100,192)[t]{ 100}
\Text(150,192)[t]{ 150}
\Text(200,192)[t]{ 200}
\Text(225,200)[l]{MeV}
\end{picture}
\end{center}\end{figure}
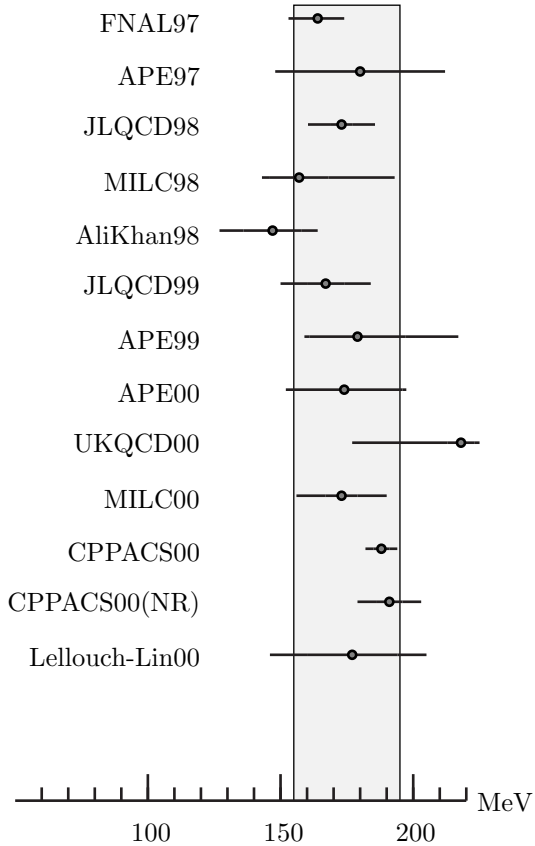

In fig.~\ref{fig:fbdiagram} I show the quark-flow diagram for the
leptonic decay of the $B$-meson. The non-perturbative QCD effects are contained in
the matrix element
\begin{equation}
\langle\,0\,|\,A_\mu(0)\,|\,B(p)\,\rangle=i f_B p_\mu\,,
\end{equation}
where $A_\mu$ is the axial-vector current with the appropriate
flavour quantum numbers. Using Lorentz and Parity Invariance we
see that all the nonperturbative QCD effects are parametrized in
terms of a single number $f_B$ the (leptonic) decay constant of
the $B$-meson~\footnote{The convention for the normalization used
here corresponds to $f_\pi\simeq 132$\, MeV.}. Quenched
calculations of $f_B$ have been performed for about 15 years now
and a careful analysis of all the systematic errors (apart from
quenching) is possible (see for example C.Bernard's review at
Lattice 2000~\cite{cblat2000}). In fig.~\ref{fig:fbquenched} I
update Claude Bernard's compilation of recent results using a
variety of different formulations for the $b$
quark~\cite{cblat2000}. His conclusion from these results for the
quenched value of $f_B$ (with which I concur) is:
\begin{equation}\label{eq:fbquenched}
f_{B\textrm{,quenched}}=175\pm 20\,\textrm{MeV}\ ,
\end{equation}
and this is shown as the shaded box in fig.~\ref{fig:fbdiagram}.

It may appear a little strange that with so many results being
included in fig.~\ref{fig:fbquenched}, the final error is as large
as in eq.~(\ref{eq:fbquenched}) and as indicated in the shaded
region of the figure. This is a manifestation of the difficulty in
controlling the systematic uncertainties, many of which are common
to the different determinations and the estimate of the overall
error requires a careful analysis of the treatment of these
uncertainties by each group . For example, we know that quenching
induces errors of $O(10\%)$ in many physical quantities. In
particular the value of the lattice spacing determined by using
different physical quantities to set the scale (e.g. $m_\rho$ or
$f_\pi$) typically also varies by this amount. Thus there is an
irreducible error in the value of $f_B$ in the quenched
approximation of about 10\% (or about 20\,MeV).

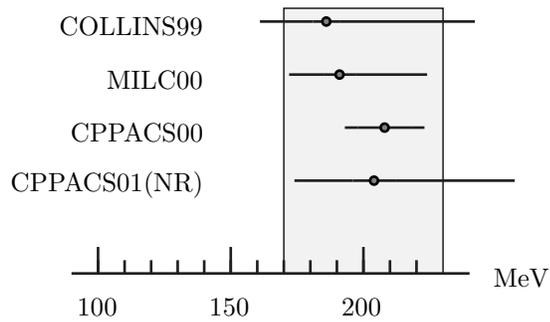
\begin{figure}[ht]
\caption{Results for $f_B$ from 4 simulations using two flavours
of sea quarks~\protect\cite{fbfiguq}. Statistical and systematic
errors have been combined in quadrature.\label{fig:fbnf2}}
\begin{center}
\begin{picture}(200,160)(70,350)
\GBox(170,400)(230,500){0.95}
\Text(140,495)[r]{COLLINS99}
\Text(140,475)[r]{MILC00}
\Text(140,455)[r]{CPPACS00}
\Text(140,435)[r]{CPPACS01(NR)}
\SetWidth{1}
\Line(186,495)(191,495)\Line(186,495)(181,495)
\GCirc(186,495){1.5}{0.5}
\Line(191,475)(197,475)\Line(191,475)(185,475)
\GCirc(191,475){1.5}{0.5}
\Line(208,455)(218,455)\Line(208,455)(198,455)
\GCirc(208,455){1.5}{0.5}
\Line(204,435)(212,435)\Line(204,435)(196,435)
\GCirc(204,435){1.5}{0.5}
\Line(191,495)(242,495)\Line(181,495)(161,495)
\Line(197,475)(224,475)\Line(185,475)(172,475)
\Line(218,455)(223,455)\Line(198,455)(193,455)
\Line(212,435)(257,435)\Line(196,435)(174,435)
\Line(90,400)(240,400)\Line(100,400)(100,410)
\Line(150,400)(150,410)\Line(200,400)(200,410)
\Line(100,400)(100,405)\Line(110,400)(110,405)
\Line(120,400)(120,405)\Line(130,400)(130,405)
\Line(140,400)(140,405)\Line(150,400)(150,405)
\Line(160,400)(160,405)\Line(170,400)(170,405)
\Line(180,400)(180,405)\Line(190,400)(190,405)
\Line(200,400)(200,405)\Line(210,400)(210,405)
\Line(220,400)(220,405)
\Text(100,392)[t]{100} \Text(150,392)[t]{150}
\Text(200,392)[t]{200}
\Text(250,400)[l]{MeV}
\end{picture}
\end{center}\end{figure}

The emphasis is now turning to unquenched calculations, see
fig.\ref{fig:fbnf2}. There is some belief that $f_{B,N_f=2}$ is
10-15\% larger that the decay constant in the quenched
approximation and C.~Bernard's conclusion is that
\begin{equation}
f_B=200\pm 30\,\textrm{MeV}\ .
\end{equation}

An important parameter which appears in analyses of the Unitarity
Triangle is
\begin{equation}
\xi\equiv\frac{f_{B_s}\sqrt{B_{B_s}}}{f_{B_d}\sqrt{B_{B_d}}}\ ,
\end{equation}
where the $B$-parameters of neutral $B$-meson mixing are defined
in section~\ref{subsec:bbbarmixing} below. It will be shown below
that lattice simulations indicate that the $B$-parameter varies
slowly with the light-quark mass and so it is interesting and
instructive to consider the ratio $f_{B_s}/f_{B_d}$. The lattice
results for this quantity have been very stable  and C.~Bernard
concludes from these results that
\begin{eqnarray}
\left(\frac{f_{B_s}}{f_B}\right)_{\textrm{quenched}}&=&1.15\pm
0.04\label{eq:fbsoverfbq}\\
\frac{f_{B_s}}{f_B}&=&1.16\pm 0.04\ .\label{eq:fbsoverfb}
\end{eqnarray}
in the quenched theory and in full QCD respectively.

Thus the ratio in eqs.~(\ref{eq:fbsoverfbq}) and
(\ref{eq:fbsoverfb}) is determined rather precisely. This should
not be too much of a surprise since the key point to note is that
for both $\xi$ and $f_{B_s}/f_{B_d}$ it is the difference from 1
which is being computed. Lattice errors of 30\% (which is a
conservative estimate) therefore correspond to errors of only
about 5\% on $\xi$.

In this short lecture I do not have the opportunity of explaining
the formulations of heavy quarks on the lattice in any detail, but
let me mention briefly the reason why different groups use
different formulations. The number of lattice points is limited by
the available computing resources and we require the volume of the
lattice to be larger than the hadrons being studied. We also
require the lattice spacing $a$ to be small enough to avoid
discretisation errors (i.e. artefacts due to the granularity of
the lattice) so that the choice of the value of $a$ is a
compromise between two sources of possible error. Typically, in
current simulations, one takes $a^{-1}\sim 2$\,-\,3\,GeV. This is
less than the mass of the $b$-quark, $m_b$, so that we cannot
study the propagation of a physical $b$-quark on presently
available lattices. To circumvent this difficulty, lattice results
for $b$-physics are obtained by taking those obtained with the
heavy-quark mass, $m_Q$, in the region of the mass of the charm
quark ($m_Q\sim m_c$) and performing the extrapolation $m_Q\to
m_b$ or by performing simulations in effective theories, such as
the {\it Heavy Quark Effective Theory} or {\it Non-Relativistic
QCD} or by a combination of these two approaches. There is
therefore reasonable control over this source of systematic
uncertainty and as computing resources increase we will be able to
verify explicitly that this is indeed the case.

\subsection{$B$-$\bar B$ Mixing}
\label{subsec:bbbarmixing}

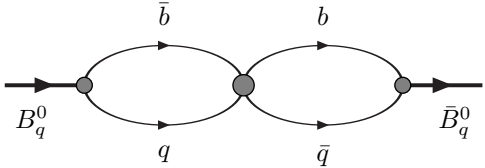
\begin{figure}
\caption{Schematic representation of $B^0$-$\bar B^0$
mixing.\label{fig:bbarmixing}}
\begin{center}
\begin{picture}(180,65)(-90,-30)
\Oval(-30,0)(15,30)(0)\Oval(30,0)(15,30)(0)
\SetWidth{1.5}\ArrowLine(-90,0)(-60,0) \ArrowLine(60,0)(90,0)
\SetWidth{0.5}
\ArrowLine(-30.5,15)(-29.5,15)\ArrowLine(-30.5,-15)(-29.5,-15)
\ArrowLine(29.5,15)(30.5,15)\ArrowLine(29.5,-15)(30.5,-15)
\GCirc(-60,0){3}{0.5}\GCirc(60,0){3}{0.5} \GCirc(0,0){4}{0.5}
\Text(-80,-8)[t]{$B_q^0$}\Text(80,-8)[t]{$\bar B_q^0$}
\Text(-30,23)[b]{$\bar b$}\Text(30,23)[b]{$b$}
\Text(-30,-23)[t]{$q$}\Text(30,-23)[t]{$\bar q$}
\end{picture}
\end{center}
\end{figure}
The quark flow diagram for $B$-$\bar B$ mixing is drawn in
fig.\ref{fig:bbarmixing}. The non-perturbative QCD effects in this
important process are contained in the single matrix element:
\begin{equation}
M(\mu)\equiv\langle\,\bar B_q^0\,|\,\bar b\gamma_\mu(1-\gamma_5)q\
\bar b\gamma^\mu(1-\gamma_5)q\,|\,B_q^0\,\rangle\,,
\end{equation}
where $q$ represents the $d$ or $s$ quark.
It is convenient and conventional to define $B$-parameters by
factorizing the vacuum saturation contribution:
\begin{equation}
 M(\mu)=\frac{8}{3}f_{B_q}^2m_{B_q}^2B_{B_q}(\mu).
\end{equation}
The $B_{B_q}(\mu)$ parameters are renormalization scheme and
scale-dependent and therefore it is again convenient and
conventional to define scheme-independent (up to NLO) quantities
\begin{equation}
\hat B_{B_q}^{nlo}=\alpha_s(\mu)^{2/\beta_0}\left[
1\,+\,\frac{\alpha_s(\mu)}{4\pi}\,J_{n_f}\right]\ B_{B_q}(\mu),
\end{equation}
where $J_{n_f}$ is a known constant calculated in perturbation
theory.

Little has changed since last year, when C.~Bernard's summary of
the corresponding results at the lattice conference
was~\cite{cblat2000}:
\begin{eqnarray}
\hat B_{B_d}&=& 1.30\pm 0.12\pm 0.13\\
f_{B_d}\sqrt{\hat B_{B_d}}&=&230\pm 40\,\textrm{MeV}\\
\frac{\hat B_{B_s}}{\hat B_{B_d}} &=&1.00\pm 0.04\\
\xi &=& 1.16\pm 0.05\ .
\end{eqnarray}

\subsection{$K^0$-$\bar K^0$ Mixing
and $B_K$}\label{subsec:bk}

The quark flow diagram is similar to that for $B_B$ above (see
fig.~\ref{fig:bbarmixing}). The non-perturbative QCD effects are
contained in the matrix element:
\begin{equation}
\langle\bar K^0|(\bar s\gamma_\mu(1-\gamma_5)d)(\bar
s\gamma_\mu(1-\gamma_5)d)|K^0\rangle\,.\label{eq:bk}
\end{equation}
Chiral symmetry plays a central r\^ole in the determination of the
corresponding $B$-parameter $B_K$. This presents a relative
difficulty for simulations performed using the Wilson formulation
of lattice fermions (or extensions of this formulation), since in
this case we don't have explicit chiral symmetry until the
extrapolation to zero quark mass is performed. Specifically, the
operator in the matrix element of eq.~(\ref{eq:bk}) mixes with
other operators of dimension 6 and the matrix elements of these
operators have to be subtracted, leading to a loss of precision.
Over the years however, techniques have been developed to perform
these subtractions non-perturbatively and reasonably effectively.

There are two recent and related proposals to circumvent the need
for the subtraction of the additional operators, based on Bernard's
observation that CPS-sy\-mm\-etry (where S=$s\leftrightarrow d$,
the interchange of the $s$ and $d$ quarks) implies that the parity-odd
component of the $\Delta S=2$ operator in eq.~(\ref{eq:bk}) renormalizes
multiplicatively. However, it is the parity-even component
which is non-zero in eq.~(\ref{eq:bk}).

The first proposal is to use \textit{twisted-mass}
QCD~\cite{twisted}:
\begin{equation}
{\cal
L}=\bar\psi(D_W+m_0+i\mu_0\gamma_5\tau^3)\,\psi+\bar s(D_W+m_0^s)s
\end{equation}
where $\psi$ represents the isodoublet of light-quarks ($\tau^3$
is a matrix in this space), $\mu_0$ is a parameter and $D_W$ is
the Wilson formulation of the Dirac operator. The
(multiplicatively renormalized) parity-odd operator in this theory
corresponds to the physical (parity-even) operator in QCD.

It is also possible to use a (chiral) Ward Identity to determine
the physical matrix element of the parity-even $\Delta S=2$
operator from a measurement of that of the parity-odd component,
$O^{\Delta S=2}_{VA}$, without the twisted
mass-term~\cite{damir2000}. A recent simulation with this
formulation gives:
\begin{equation}
B_K(2\,\textrm{GeV})=0.73\pm 0.07^{+0.05}_{-0.01}.
\end{equation}

Nevertheless, the most precise evaluation of $B_K$ comes from
simulations using the staggered formulation of QCD, in which
chiral symmetry is explicit at the expense of a more complicated
flavour structure. The evaluation of $B_K$ is also an excellent
testing ground for new formulations of chiral fermions based on
the Ginsparg-Wilson relation and results of $\hat B_K=0.787(8)$
and $0.737(11)$ have been reported by the CP-PACS~\cite{cppacsbk}
and RBC~\cite{rbcbk} collaborations using Domain Wall Fermions.

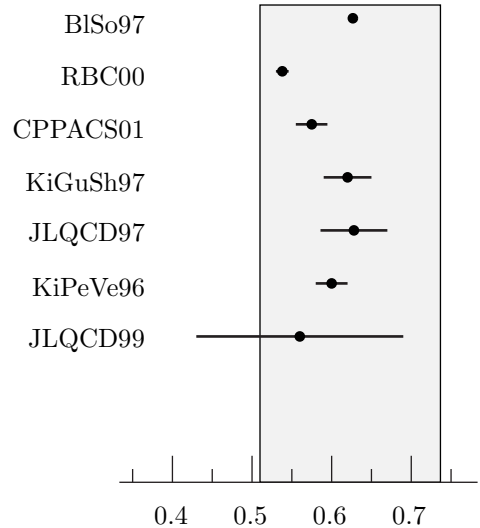
\begin{figure}
\caption{Results for $B_K^{\textrm{NDR}}(2\,\textrm{GeV})$ in the
continuum limit from various groups (mainly) in the quenched
approximation~\protect\cite{bkresults}. Statistical and systematic
errors have been combined in quadrature (where these have been
presented).\label{fig:bkresults}}
\begin{center}
\begin{picture}(200,210)(40,290)
\GBox(153,320)(221,500){0.95} \Text(110,495)[r]{BlSo97}
\Text(110,475)[r]{RBC00} \Text(110,455)[r]{CPPACS01}
\Text(110,435)[r]{KiGuSh97} \Text(110,415)[r]{JLQCD97}
\Text(110,395)[r]{KiPeVe96} \Text(110,375)[r]{JLQCD99}
\SetWidth{1} \Line(188,495)(187,495)\Line(188,495)(189,495)
\GCirc(188,495){1.5}{0}
\Line(161.4,475)(163.8,475)\Line(161.4,475)(159,475)
\GCirc(161.4,475){1.5}{0}
\Line(172.5,455)(178.4,455)\Line(172.5,455)(166.5,455)
\GCirc(172.5,455){1.5}{0}
\Line(186,435)(177,435)\Line(186,435)(195,435)
\GCirc(186,435){1.5}{0}
\Line(188.4,415)(201,415)\Line(188.4,415)(175.8,415)
\GCirc(188.4,415){1.5}{0}
\Line(180,395)(186,395)\Line(180,395)(174,395)
\GCirc(180,395){1.5}{0}
\Line(168,375)(207,375)\Line(168,375)(129,375)
\GCirc(168,375){1.5}{0} \SetWidth{0.5}
\Line(100,320)(235,320)\Line(120,320)(120,330)
\Line(150,320)(150,330)\Line(180,320)(180,330)
\Line(210,320)(210,330)
\Line(135,320)(135,325)\Line(165,320)(165,325)
\Line(195,320)(195,325)\Line(225,320)(225,325)
\Line(105,320)(105,325)
\Text(120,312)[t]{0.4} \Text(150,312)[t]{0.5}
\Text(180,312)[t]{0.6} \Text(210,312)[t]{0.7}
\end{picture}\end{center}\end{figure}

Fig.~\ref{fig:bkresults} contains a compendium of recent results
for $B_K$ compiled by L.Lellouch, the rapporteur at 2000 Lattice
conference, whose conclusion from these results
is~\cite{lllat2000}:
\begin{eqnarray}
B_K^{\textrm{NDR}}(2\,\textrm{GeV})&=&0.628\pm0.042\pm 0.099\nonumber\\
\to \hat{B}_K^{\textrm{NLO}}&=&0.86\pm0.06\pm0.14\ .
\end{eqnarray}
The errors include estimates of those due to quenching obtained
using Sharpe's analysis with quenched chiral perturbation
theory~\cite{ssqcpt}.

In extensions of the standard model, and in particular in
supersymmetric theories, $\Delta S=2$ operators other than that in
eq.~(\ref{eq:bk}) also contribute to $B_K$. These can also be
evaluated in lattice simulations~\cite{bsmbk1,bsmbk2} thus helping to
provide constraints on the properties of supersymmetric models.

\section{$K\to\pi\pi$ Decays}
\label{sec:kpipi} The evaluation of the amplitudes for nonleptonic
weak decays, particularly of $B$ and $K$ mesons, represents a
major challenge for lattice physicists.  At this conference we
have heard many interesting experimental results for two-body
exclusive decays of $B$-mesons such as $B\to\pi\pi$ or $B\to\pi K$
decays~\cite{bdecays}, but, at present, we are unable to perform
lattice calculations of the corresponding matrix elements.
Considerable progress is being made however, towards reliable
calculations of $K\to\pi\pi$ decays, including quantitative
studies of the $\Delta I=1/2$ rule and an evaluation of
$\varepsilon^\prime/\varepsilon$. In this section I will outline
some of this progress and attempt to convey the optimism we feel
for future prospects. For a more comprehensive and detailed recent
review see ref.~\cite{gmkaon01} and references therein.

The lattice contribution to the evaluation of $K\to\pi\pi$
amplitudes begins with the use of the operator product expansion
leading to an expression for the $\Delta S=1$ effective weak
Hamiltonian in terms of Wilson coefficient functions $C_i$ and
renormalized local operators $\hat O_i(\mu)$:
\begin{equation}
{\cal H}_W^{\Delta S=1}=-\frac{G_F}{\sqrt 2}\sum_i C_i(\mu) \hat
O_i(\mu)\ . \label{eq:hweff}\end{equation}
The $C_i$'s contain the perturbative QCD effects which give the
evolution from the mass of the $W$ to the (perturbative)
renormalization scale $\mu$. The non-perturbative physics is
contained in the matrix elements $\langle\pi\pi|\hat
O_i|K\rangle$, and the r\^ole of lattice simulations is to
evaluate these matrix elements.

From lattice computations one obtains the matrix elements of the
bare lattice operators with the lattice spacing $a$ as the
ultra-violet cut-off. From these we must construct the finite
matrix elements of renormalized operators, and this
\textit{ultra-violet problem} is, in principle at least, fully
solved~\cite{uv}. Several non-perturbative techniques  have been
developed to determine the corresponding renormalization
coefficients~\cite{npr1,npr2,npr3} (for recently calculated
renormalization constants in perturbation theory for the Domain
Wall formulation of lattice fermions see ref.~\cite{uvdomain}).

Two main approaches are used to determine the decay amplitudes
from lattice simulations:\\ i) The $K\to\pi$ (and $K\to$ vacuum)
matrix elements are computed directly, and the $K\to\pi\pi$ matrix
elements are obtained using soft-pion theorems and (lowest order)
chiral perturbation theory.\\ ii) The $K\to\pi\pi$ matrix elements
are computed directly.\\ Although, at first sight, it may appear
that the second approach is clearly better (and I expect that it
will eventually become the standard one), it does involve a
two-hadron final state which presents some subtleties~\cite{mt}.
We are trying to determine physical decay amplitudes from matrix
elements computed in a finite Euclidean volume. Lellouch and
L\"uscher have initiated substantial progress towards the solution
of this {\it infrared} problem and I will briefly review this in
section~\ref{subsec:fv}. In section~\ref{subsec:prospects} I will
review some recent numerical results and discuss prospects for
future calculations of $K\to\pi\pi$ amplitudes.

\subsection{$K\to\pi\pi$ Decays in a Finite Volume}
\label{subsec:fv}

The \textit{infrared} problem for the evaluation of $K\to\pi\pi$
decay amplitudes arises from two sources, the unavoidable
continuation of the theory to Euclidean space-time (the
Maiani-Testa Theorem~\cite{mt}) and the use of a finite volume in
numerical simulations. An important step towards the solution of
this problem has been achieved by Lellouch and L\"uscher~\cite{ll}
who derived a relation between the $K\to\pi\pi$ matrix element in
a finite volume and the physical decay amplitude:
\begin{eqnarray}
|\langle\,\pi\pi\,|\,{\cal H}_W(0)\,|\,K\,\rangle|^2&=&
|_V\langle\,\pi\pi\,|\,{\cal
H}_W(0)\,|\,K\,\rangle_V|^2\nonumber\\ &&\hspace{-1.3in} \times
8\pi V^2\left(\frac{m_K}{k_\pi}\right)^3\,
\{q\phi^\prime(q)+k\delta_0^\prime(k)\}_{k=k_\pi}.
\label{eq:ll}
\end{eqnarray}
In the first line of eq.~(\ref{eq:ll}), the left-hand side is the
infinite-volume matrix elements and the right-hand side is the
finite-volume (which might, for example, be computed in a lattice
simulation). The second line is the factor relating these and the
main message that I am trying to convey here is that there is a
known factor which relates the physical amplitude and the
finite-volume matrix element. In eq.~(\ref{eq:ll}) $k$ is related
to the centre of mass energy $W$ by:
\begin{equation}
W=2\sqrt{m_\pi^2+k^2}\ \textrm{and}\
k_\pi=\frac12\sqrt{m_K^2-4m_\pi^2}\ ;
\end{equation} $q=kL/2\pi$ where $L$ is
the length of the lattice, $\phi(q)$ is a known kinematic function
of $q$ and is a consequence of the cubic shape of the finite
lattice and $\delta(k)$ is the physical (infinite-volume) s-wave
phase-shift (the explicit formula in eq.~(\ref{eq:ll}) applies
only to this partial wave). The remaining finite-volume
corrections decrease exponentially as the volume increases.

In lattice simulations we calculate correlation functions at large
(Euclidean) times so as to isolate the ground state. Most
frequently, as for example in the computations used to obtain the
results presented in section~\ref{sec:ckm}, the interpolating
operators used in the correlation functions are such that the
ground state corresponds to the lightest particle with some
specified quantum numbers. For $K\to\pi\pi$ decays we are also
interested in two-particle states. One of the consequences of the
Maini-Testa theorem~\cite{mt} is that such correlation functions
are dominated at large times by the states in which the kaon and
each of the two-pions are (almost) at rest. Such a kinematical
situation is clearly unphysical. Lellouch and L\"uscher~\cite{ll}
make the interesting and significant observation that, since
energy levels in a finite box are discrete, it is possible to tune
the size of the box (lattice) in such a way that the energy of the
first excited two-pion state is precisely the mass of the kaon.
This would require a volume of about 6\,fm, somewhat larger than
currently used, but one which should become accessible in quenched
simulations with the next generation of dedicated computers.
Although the first excited state is more difficult to extract than
the ground state, this is not likely to present a major
difficulty.

Lellouch and L\"uscher derive the formula in eq.~(\ref{eq:ll}) for
a lattice with a fixed large volume $V$, chosen in such a way that
the decay of the kaon with physical kinematics corresponds to one
of the two-pion energy levels accessible on this lattice (such as
the first excited state mentioned above). We have recently
rederived eq.~(\ref{eq:ll}), taking the $V\to\infty$ limit at
fixed \textit{physics} starting from the L\"uscher quantization
condition~\cite{ml}
\begin{equation}
\phi(q)+\delta_0(k)=n\pi\,,
\end{equation}
which gives the spectrum of two-particle states in a finite cubic
volume~\cite{lmst}. We were able to establish the validity of
eq.~(\ref{eq:ll}) for all elastic states below the inelastic
threshold, with exponential accuracy in the volume, extending the
derivation in ref.\cite{ll} which was presented for the lowest
seven levels. We were also able to demonstrate that the formula
was valid when the two-pion energy does not match the mass of the
kaon $m_K$ and the inserted operator (e.g. one of the $\hat O_i$
in the operator product expansion for the weak Hamiltonian
eq.~(\ref{eq:hweff})\,) carries non-zero energy and momentum. This
is particularly useful in attempts to determine the coefficients
of the operators appearing in higher orders of the chiral
expansion (see sec.~\ref{subsec:prospects} below).

\subsection{Status and Prospects for the Evaluation of
$K\to\pi\pi$ Decays.}
\label{subsec:prospects}

In this section I present some recent lattice results of the
matrix elements which contribute to $K\to\pi\pi$ decay amplitudes
and discuss the exciting prospects for further progress in
improving the precision. Since the delivery of this lecture in
July, two groups have presented results which include the
evaluation of $\Delta I=1/2$ amplitudes including the
determination of $\varepsilon^\prime/\varepsilon$, and in view of
the interest in such calculations I will also comment briefly on
these results.

\begin{figure}[t]
\caption{Two contributions to nonleptonic kaon decay amplitudes.
\label{fig:kaonamps}}
\begin{center}
\begin{picture}(170,120)(-80,-50)
\SetWidth{1.5}
\ArrowLine(-80,0)(-53,0)\ArrowLine(53,0)(80,0)
\ArrowLine(43.64,46)(70.6,46)
\SetWidth{0.5}
\Oval(0,0)(20,50)(0)\Oval(23.3,36)(6,20)(30)
\GCirc(0,20){2.5}{0.5}
\GCirc(6,26){2.5}{0.5}
\GCirc(-50,0){3}{0.5}
\GCirc(50,0){3}{0.5}\GCirc(40.64,46){3}{0.5}
\Text(-70,-8)[t]{$K$}\Text(85,0)[l]{$\pi$}
\Text(77,46)[l]{$\pi$}\Text(-30,22)[b]{$s$}
\Text(32,18)[b]{$u$}
\Text(0,-45)[t]{a) Disconnected Emission}
\end{picture}
\begin{picture}(170,180)(170,-50)
\SetWidth{1.5}
\ArrowLine(170,30)(200,30)\ArrowLine(303,80)(330,80)
\ArrowLine(303,-20)(330,-20)
\SetWidth{0.5}
\Oval(250,80)(15,10)(0)
\Line(200,30)(300,80)\Line(200,30)(300,-20)
\Line(270,30)(300,80)\Line(270,30)(300,-20)
\GCirc(250,55){2.5}{0.2}
\GCirc(250,65){2.5}{0.2}
\GCirc(200,30){3}{0.5}
\GCirc(300,80){3}{0.5}\GCirc(300,-20){3}{0.5}
\GCirc(270,30){1}{0.8}
\Text(180,22)[t]{$K$}\Text(335,80)[l]{ $\pi$} \Text(335,-20)[l]{
$\pi$}\Text(220,47)[b]{ $s$} \Text(270,73)[b]{ $d$}
\Text(250,-45)[t]{b) Disconnected Penguin}
\end{picture}\end{center}
\end{figure}
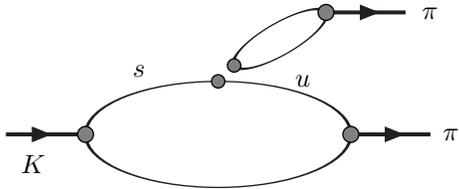

A number of different mechanisms contribute to the amplitudes for
$K\to\pi\pi$ decays, and in fig.~\ref{fig:kaonamps} I show some
examples. The four-quark vertex at which there is an $s\to u$ or
$s\to d$ transition represents the insertion of one of the
four-quark operators appearing in the effective Hamiltonian in
eq.~(\ref{eq:hweff}). Lattice calculations have shown that it is
not possible to explain the $\Delta I=1/2$ rule, i.e. the
experimentally observed enhancement of the amplitudes for $\Delta
I$=1/2 decays by a factor of about 22 relative to those for
$\Delta I$=3/2 decays, with emission diagrams only. I start with a
discussion of attempts to evaluate the penguin contractions using
Wilson-like lattice fermions. In order to obtain the physical
contribution from the penguin diagrams, in general we have to
subtract large unphysical artefacts, terms which diverge as
inverse powers of the lattice spacing (power divergences). From
the diagram in fig.~\ref{fig:kaonamps}\,(b) we can see that the
inserted four-quark operator can mix with quark bilinears of the
type $\bar d\Gamma s$ (where $\Gamma$ is one of the Dirac
matrices) unless there are symmetries to prevent this mixing.
Although lattice symmetries do soften the divergences
corresponding to the mixing, large subtractions (direct or
indirect) are nevertheless unavoidable.  This is the reason for
the absence up to now of sufficiently precise results for $\Delta
I=1/2$ decays.

\begin{figure}[ht]\caption{Raw lattice data for the matrix
element $_{I=0}\langle\pi\pi|O^-|K\rangle$ as a function of the
time.\label{fig:om}}
\vspace{-25pt}
\includegraphics[width=0.9\hsize,angle=270]{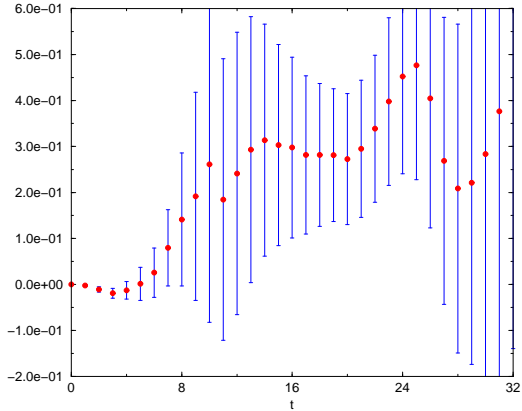}
\end{figure}

The excitement in the \spqr (South\-amp\-ton-Rome-(QCD)-Paris)
collaboration, of which I am a member, is due to the fact that for
the first time we have a lattice signal for the amplitudes, which
we will analyse to determine $\varepsilon^\prime/\varepsilon$ and
study the $\Delta I=1/2$ rule. This is partly due to improved
theoretical techniques to reduce  the subtractions and to deal
with them non-perturbatively, and partly due to improved computing
facilities. In fig.~\ref{fig:om} I show the (preliminary) raw
lattice data for the ratio of correlation functions as a function
of the \textit{time} $t$, from which one of the relevant matrix
elements, $_{\pi\pi}\langle\pi\pi|O_-|K\rangle$, is determined.
There is a stable region in $t$, where the matrix element can be
seen to be non-zero and we will increase the statistics to reduce
the error which is currently still large.

\begin{figure}\begin{center}
\caption{An example of the very large subtractions present in some
lattice evaluations of the $\Delta I=1/2$ transitions. The figure
shows the values of the unsubtracted and subtracted bare matrix
elements $_{I=0}\langle\,\pi\pi|O_6\,|K\rangle$ as a function of
the quark mass from a simulation by the CP-PACS
Collaboration~\protect\cite{cppacs}.\label{fig:o6subs}}
\includegraphics[width=\hsize]{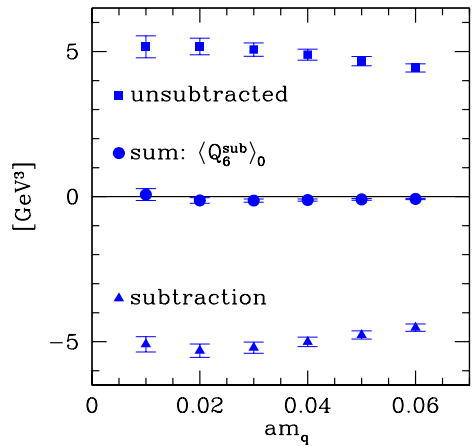}
\vspace{-0.5in}
\end{center}\end{figure}

In fig.~\ref{fig:o6subs} I illustrate the huge subtractions which
generally have to be performed with data from the CP-PACS
collaboration for the matrix element of $O_6$, which is one of the
important operators contributing to
$\varepsilon^\prime/\varepsilon$.

\begin{figure}[t]
\begin{center}
\caption{Preliminary data for the matrix element of the
electroweak penguin operator $O_8$ for a particular choice of the
quark mass as a function of the lattice time
$t$~\protect\cite{spqrlatt01}.\label{fig:spqro8}}\vspace{-0.2in}
\includegraphics[width=0.85\hsize,angle=270]{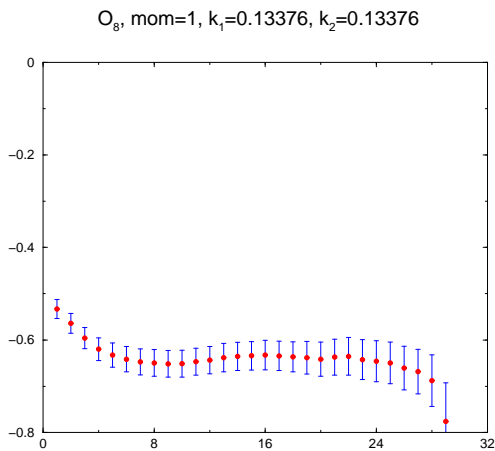}
\end{center}\end{figure}

I now turn to the matrix elements of $\Delta I=3/2$ operators
between the kaon and two-pion states. These can can now be
evaluated with good precision, and as an example I present in
fig.~\ref{fig:spqro8} the $\Delta I=3/2$ matrix elements of the
electroweak operator $O_8$ which is one of the key components in
the evaluation of $\varepsilon^\prime/\varepsilon$.  We are
currently undertaking a detailed study, up to next-to-leading
order in the chiral expansion for the $\Delta I=3/2$ matrix
elements of these operators ($O_{4,7,8}$). Preliminary results
indicate that it is possible to obtain a $\Delta
I=3/2\,K\to\pi\pi$ decay amplitude, which is nevertheless
consistent with a large $B_K$ (see sec.\ref{subsec:bk}).

There is a long history of lattice studies of kaon decays in which
operator matrix elements of the type $\langle \pi|O|K\rangle$ are
computed and combined with soft-pion theorems and chiral
perturbation theory to obtain the decay amplitudes. Here I mention
one such recent result, for the matrix element of the electroweak
penguin operator $O_8$~\cite{bsmbk2}, for which lattice results
give significantly smaller values than other determinations, e.g.
\begin{equation}
|_{I=2}\langle\pi\pi|O_8|K^0\rangle|=(0.5\pm 0.1)\,\textrm{GeV}^3\
, \label{eq:o8latt}\end{equation} in the NDR renormalization
scheme at a scale of 2\,GeV (such a small result was also
confirmed from computations of $K\to\pi\pi$ amplitudes by the
\spqr collaboration in the presentation of their preliminary data
at Lattice 2001~\cite{spqrlatt01}). This can be compared to the
larger values, e.g. 2.22$\pm$0.67\,GeV$^3$~\cite{dg} and
(3.5$\pm$1.1)\,GeV$^3$~\cite{kpd} obtained using other (continuum)
techniques such as dispersion relations and the $1/N_c$
expansion~\footnote{After including the recently calculated
$O(\alpha_s^2)$ corrections~\cite{dg} the authors of
ref.~\cite{kpd} find their results becomes $(2.4\pm
0.8)$\,Gev$^3$~\cite{kpd2}.}. There is also a recent continuum
result of 1.2$\pm$0.5\,Gev$^3$~\cite{bgp} obtained using spectral
functions. A large value of the matrix element of $O_8$ would
require very large values indeed for the matrix elements of $O_6$
in order to explain the measured value of
$\varepsilon^\prime/\varepsilon$ and so it will be interesting to
follow the developments in these calculations.

As mentioned above, two groups have recently presented results for
$\varepsilon^\prime/\varepsilon$ and other properties of
$K\to\pi\pi$ decay amplitudes, from a computation of $K\to\pi$
matrix elements using the Domain Wall Fermion formulation for
lattice quarks (these results were presented since this lecture
was delivered). In view of the experimental
result~\cite{expepsprime} $\varepsilon^\prime/\varepsilon=(17.2\pm
1.8)\,10^{-4}$ these results may seem somewhat disturbing:
\begin{eqnarray}
\varepsilon^\prime/\varepsilon&=& (-8 \div -4)\,10^{-4}\ \
\textrm{RBC}\mbox{\,\cite{rbc}}\label{eq:rbcepe}\\
\varepsilon^\prime/\varepsilon&=& (-7 \div -2)\,10^{-4}\ \
\textrm{CP-PACS}\mbox{\,\cite{cppacs}}. \label{eq:cppacsepe}
\end{eqnarray}
Of course, it would be very exciting to be able to confidently
deduce the existence of new physics from the discrepancy between
the lattice results in eqs.~(\ref{eq:rbcepe}) and
(\ref{eq:cppacsepe}) and the experimentally measured value of
$\varepsilon^\prime/\varepsilon$. Such a conclusion remains a
tantalizing possibility. Before this can be done however, we need
to be reassured that the systematics of the computations are
sufficiently under control. Although the results for
$\varepsilon^\prime/\varepsilon$ from refs.\cite{rbc} and
\cite{cppacs} are consistent with each other, this is not the case
for some other quantities, including some of the separate
components in $\varepsilon^\prime/\varepsilon$. For example both
groups find a large, but different, value for the ratio
Re$(A_0)/$Re$(A_2)$, where 0 and 2 denote the isospin of the
two-pion system (and hence an enhancement of $\Delta I=1/2$
decays\,!). The RBC collaboration find $\textrm{Re}
A_0/\textrm{Re} A_2=24\div 27$ (remarkably close to the
experimental value) whilst CP-PACS find $\textrm{Re}
A_0/\textrm{Re} A_2=9\div 12$. Since both groups use similar
methods (but with some important differences, for example in the
normalisation of the operators) we need to understand the reason
for discrepancies such as these.

It has to be stressed that this calculation is much more difficult
and subtle than those reported in section~\ref{sec:ckm}. Given the
huge subtraction of power divergences, illustrated in
fig.~\ref{fig:o6subs}, a good understanding of the chiral
properties of the theory with the parameters used in the
simulation is crucial. The computations were performed using
Domain Wall Fermions with $N_5=16$ points (where $N_5$ is the
number of points in the fifth direction), and although both groups
claim that this is sufficient for the residual chiral symmetry
breaking effects to be fully under control, it would be very
reassuring to confirm this by increasing $N_5$ whilst keeping the
other parameters fixed. We also need to be able to understand the
consequences for these calculations of the recent observation by
Golterman and Pallante~\cite{gp} that in the quenched
approximation there are additional (spurious) chiral logarithms.
It should also be remembered that the results for $K\to\pi\pi$
amplitudes in eqs.~eqs.~(\ref{eq:rbcepe}) and (\ref{eq:cppacsepe})
were obtained from the determination of $K\to\pi$ matrix elements
using lowest order chiral perturbation theory, and one can ask
about the precision of this procedure. Nevertheless, in spite of
caveats such as these, these new results are very exciting and
mark the beginning of a new era in lattice studies of kaon decays.

\section{Conclusion}
\label{sec:concs}
Lattice simulations provide the opportunity to evaluate
non-perturbative QCD effects from first principles with no model
assumptions or parameters. There is a large range of quantities of
central importance to particle physics which are being computed in
lattice simulations (indeed it is far too large a range to be
considered in a review talk of 30 minutes). For some quantities,
such as those which enter into the analysis of the Unitarity
Triangle which were discussed in section~\ref{sec:ckm}, the
emphasis is now on the reduction of systematic errors. For others,
such as the evaluation of nonleptonic weak decays in general and
$K\to\pi\pi$ decays in particular, we are still learning how best
to extract the physical quantities. The range of quantities which
can be studied is constantly expanding.

\section*{Acknowledgements} I am indebted to the rapporteurs at the 2000
International Symposium on Lattice Field Theory, Claude Bernard
(Heavy Quark Physics) and Laurent Lellouch (Light-Hadron Weak
Matrix Elements), for their kind permission to reproduce plots and
compilations from their talks. I warmly thank my collaborators,
David Lin, Guido Martinelli, Mauro Papinutto  and Massimo Testa
for many stimulating discussions. The written version of this talk
was prepared during the long-term workshop on Lattice QCD and
Hadron Phenomenology at the Institute for Nuclear Theory of the
University of Seattle. I thank the organizers, Martin Golterman
and Steve Sharpe, and the participants for many stimulating
discussions on the subject of this talk. Finally I am enormously
grateful to my scientific secretary, Federico Mescia, for his
generous help.

I acknowledge support from PPARC through grants
PPA/G/O/\-1998/\-00525 and PPA/G/S/1998/00529 and from the
European Union by grant HTRN-CT-2000-00145.

\end{document}